\documentclass[12pt,showpacs,showkeys,amsmath,aps,prc]{revtex4}

\usepackage{amsfonts}
\usepackage{mathrsfs}
\usepackage{subfigure}
\usepackage{supertabular}
\usepackage{color,graphicx}
\usepackage[bookmarksopen]{hyperref}

\begin{document}

\title{Non-Singlet spin structure
function in the valon model and low x scaling behavior of
$g_{1}^{NS}$ and $g_1^p$ }

\author{Fatemeh Taghavi-Shahri}
\affiliation{School of Particles and Accelerators, Institute for
Research in Fundamental Sciences (IPM), P.O. Box 19395-5531,
Tehran, Iran\\
 Email:  f\_taghavi@ipm.ir}
\author{Firooz Arash}
\affiliation{Physics Department, Tafresh University, Tafresh,
Iran\\}

\date{\today}

\begin{abstract}
A next-to-leading order QCD calculation of non-singlet spin
structure function, $g_{1}^{NS} $ is presented within the valon
representation of Hadrons. In the valon model, it is assumed that
a nucleon is composed of three dressed valence quarks: the valons
which have their own internal structure, the valence quark with
its associated sea quarks and gluons.  The results are in good
agreement with all available data from SMC, E143, HERMES and with
the newly released data from COMPASS experiments. It appears that
the small x tail of $g_{1}^{NS}$ can be described by a single
Regge-type exchange. The relevant parameter of this exchange is
given. Finally we show that the polarized proton structure
function has a scaling behavior at small x. The relevant
parameters of this behavior are given too.
\end{abstract}

 \maketitle

%%%%%%%%%%%%%%%%%%%%
%%%%%%%%%%%%%%%%%%%%

\section{Introduction}
Deep Inelastic Scattering (DIS) of leptons from the nucleon has
served as an important tool for the investigation of the nucleon
substructure and is one of the key areas  for testing the
Quantum ChromoDynamics (QCD).\\
Spin is a fundamental properties of the nucleon and the spin
structure of nucleon has been the subject of heated debates over
the past twenty years. The key question is that how the spin of
the nucleon is distributed among its constituent partons. That is,
the  determination and understanding of the shape of quarks and
gluon spin distribution functions, $\delta q_{f}(x,Q^{2})$, have
became an important issue.
\\We utilized the valon model \cite{1}
to study the polarized nucleon structure. In the valon model, it
is assumed that a nucleon is composed of three dressed valence
quarks: the valons. Each valon has its own internal structure, the
valence quark with its associated sea quarks and gluons which can
be probed at high enough $Q^2$. At low $Q^2$, a valon behaves as a
valence quark. The valons play a role in scattering problems as
the constituents do in bound state problems. It is assumed that
the valons stand at a level in between hadrons and partons and
that the valon distributions are independent of the probe or
$Q^2$. In this representation a valon is viewed as a cluster of
its own partons. The evolution of the parton distributions in a
hadron is effected through the evolution of the valon structure,
as the higher resolution of a probe reveals the parton content of
the valon. This model has yielded  excellent results for
un-polarized structure functions, \cite{1,2,3,4,5,6}.
 It has also been applied to the polarized nucleon structure function \cite{7,8} with interesting results .\\
  In this paper we would like
to concentrate on the non-singlet part of the polarized nucleon
structure function; because of its simplicity and thus its
transparency. In addition, there are more accurate data which are
extended to fairly small x region: $x<0.01$. That makes the
comparison with the theory more meaningful. Recently COMPASS
Collaboration released data on $g_1^{NS}$ to test the Bjorken sum
rule with more accuracy \cite{9}. Therefore, in this paper we have
further attempted to show the application of using this model for
studying the nucleon structure functions.
 \\This paper is organized as follows. In section 2,
a brief general outline is presented on the calculation of the
polarized nucleon structure function in the valon model. Then we
calculate the non-singlet spin structure function in section 3.
Finally, section 4 is devoted to study the Regge behavior of
$g_{1}^{NS} $ at small x and scaling behavior of $g_{1}^{p}$.
Then, we will finish with conclusions.

\section{Valon model and polarized  hadron structure function}

The connection between bound state problem at the hadronic scale
that occurs only at low $Q^2$ and deep inelastic scattering  at
high  $Q^2$ can be investigating by introducing the valons. Each
valon is a dressed valence quark, i.e., each being a valence quark
with its associated sea quarks and gluons which can be resolved
only at high $Q^2$. At low  $Q^2$ valon behaves as a CQ because
its internal structure can not be resolved. Thus the valon
distribution in a hadron is the wave-function square of the CQs,
whose structure functions are described by PQCD at high  $Q^2$
\cite{2}.

\begin{figure}[htp]
\centerline{\begin{tabular}{cc}
\includegraphics[width=6 cm]{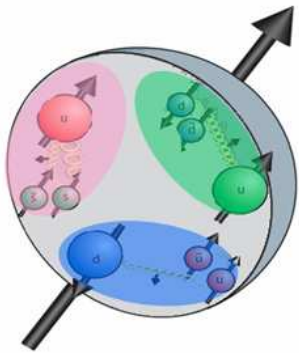}
\end{tabular}}
 \caption{\footnotesize (Color online) The schematic picture of the valon
 model.}

\end{figure}
 The valon model is essentially a two components model. In this
 framework, the structure function $F_2(x,Q^2)$ of a hadron is a
 convolution of the valon distribution $G_{\it{valon}}^{h}(y)$ and
 the structure function of the valon,
 $F^{\it{valon}}_{2}(z,Q^{2})$

\begin{equation}
F^{h}_{2}(x,Q^{2})=\sum_{\it{valon}}\int_{x}^{1}dy
G_{\it{valon}}^{h}(y) F^{\it{valon}}_{2}(\frac{x}{y},Q^{2})
\end{equation}

In a similar way the parton distribution functions in a hadron can
be obtained by

\begin{equation}
q(x,Q^{2})=\sum_{\it{valon}}\int_{x}^{1}dy G_{\it{valon}}^{h}(y)
q^{\it{valon}}(\frac{x}{y},Q^{2})
\end{equation}

where $G_{\it{valon}}^{h}(y)$ is the valon distribution inside the
hadron. It means the probability of finding a valon with momentum
fraction of $y$ in the hadron. The  description of the
$G_{\it{valon}}^{h}(y)$ is given in \cite{1,2}.

In polarized case, the helicity distributions of various partons
in a hadron in the framework of this model are given by:
\begin{equation}
\delta q_{{i}}^{\it{h}}(x,Q^2)=\sum \int_{x}^{1}
\frac{dy}{y}\delta G_{\it{valon}}^{h}(y)  \delta
q_{{i}}^{\it{valon}}(\frac{x}{y},Q^2)
\end{equation}
where $\delta G_{\it{valon}}^{h}(y)$ is the helicity distribution
of the valon in the hosting hadron (probability of finding the
polarized valon inside the polarized hadron). It is related to
unpolarized valon distribution by:
\begin{equation}
\delta G_{j}(y) = \delta F_{j}(y) G_{j}(y)
 =N_{j}y^{\alpha_{j}}(1-y)^{\beta_{j}}(1+ a_{j} y^{0.5} + b_j y +c_j y^{1.5} +d_j y^2)
\end{equation}

$G_{j}(y)$ are the unpolarized valon distributions, where $j$
refers to U and D type valons(Regrettably, the above equation was
erroneous in Ref. \cite{7} in which  $\delta G_{j}(y)$ was
replaced by $\delta F_{j}(y)$ ). Polarized valon distributions are
determined by a phenomenological argument for a number of hadrons
\cite{7,8}. We summarized the parameters for Eq.(2.4) in Table 1.

\begin{figure}[htp]
\centerline{\begin{tabular}{cc}
\includegraphics[width=8 cm]{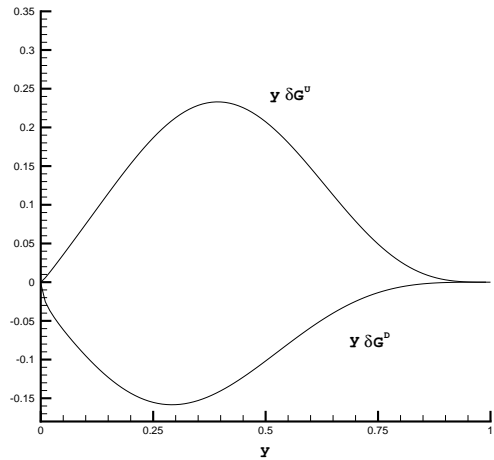}
\end{tabular}}
 \caption{\footnotesize
Polarized valon distributions for U and D valons inside the
proton.   }

\end{figure}

\begin{table}
\begin{center}
\begin{tabular}{|c|c|c|c|c|c|c|c|c|}
\hline $valon (j)$ & $N_{j}$ & $\alpha_{j}$ &
$\beta{_j} $ & $a_j $ & $b_j $ & $c_j $ & $d_j$ \\
\hline
$U$ & 3.44 & 0.33 & 3.58 & -2.47 & 5.07 & -1.859 & 2.780 \\
$D$ & -0.568 & -0.374 & 4.142 & -2.844 & 11.695 & -10.096 & 14.47
\\
\hline
\end{tabular}
\caption{Numerical values of the parameters in Eq. (4). }

\end{center}
\end{table}

We come back to Eq.(3), $\delta q_{{i}}^{\it{valon}}(z=x/y,Q^2)$
is the polarized parton distribution in the valon . Polarized
parton distributions inside the valon are evaluated according to
the DGLAP evolution equation subject to physically sensible
initial conditions.

\begin{equation}
\delta q^{NS
\pm}(n,Q^{2})=\{1+\frac{\alpha_{s}(Q^2)-\alpha_{s}(Q_{0}^{2})}{2
\pi}(\frac{-2}{\beta_{0}})(\delta {\bf{P}}^{(1)n}_{NS
\pm}-\frac{\beta_{1}}{2 \beta_{0}}\delta
{\bf{P}}^{(0)n}_{qq})\}{\bf{L}}^{-(\frac{2}{\beta_{0}})\delta
P^{(0)n}_{qq}}\delta q^{NS \pm}(n,Q_0^{2})
\end{equation}

\begin{eqnarray}
 \left( \begin{array}{c}
 \delta q^{S}(n,Q^{2}) \\ \delta g (n,Q^{2})
 \end{array} \right)
=\{{\bf{L}}^{-(\frac{2}{\beta_{0}})\delta \hat{P}^{(0)n}}+
\frac{\alpha_{s}(Q^2)}{2 \pi}{\bf{\hat{U}
L}}^{-(\frac{2}{\beta_{0}})\delta
\hat{P}^{(0)n}}-\frac{\alpha_{s}(Q_{0}^{2})}{2
\pi}L^{-(\frac{2}{\beta_{0}})\delta
\hat{P}^{(0)n}}{\bf{\hat{U}}}\} \left( \begin{array}{c}
 \delta q^{S}(n,Q_0^{2}) \\ \delta g (n,Q_0^{2})
 \end{array} \right)
\end{eqnarray}

The detail of above equations are given at \cite{10}. As presented
in  \cite{7} and \cite{8}, we have calculated the polarized
nucleon structure function in the valon model. We have worked in
$\overline{MS}$ scheme with $\Lambda_{QCD}=0.22$ $GeV$ and
$Q_{0}^{2}=0.283$ $GeV^2$. The initial motivation for this value
of $Q_{0}^{2}$ comes from the phenomenological consideration that
requires us  to choose the initial input densities as $\delta(z -
1) $ at $Q_{0}^{2}$. The valon structure function has the property
that it becomes $\delta(z - 1) $ as $Q^2$ is extrapolated to
$Q_{0}^{2}$ (beyond the region of validity). This mathematical
boundary condition means that the internal structure of the valon
can not be resolved at $Q_{0}^{2}$ in the NLO approximation. It
also means that at initial scale of $Q_{0}^{2}$, the nucleon can
be considered as a bound state of three valence quarks that carry
all the momentum and the spin of the nucleon. As $Q^2$ is
increased other partons can be resolved at the nucleon. It is also
interesting to note that this value of $Q^2$ is very close to the
transition region reported by the CLAS Collaboration. Measurement
of the first moment of the proton structure function at CLAS shows
that there is a transition region around $Q^2=0.3 GeV^2$ \cite{11}
. Therefore the initial input densities to solve the DGLAP
equations inside the valon are

\begin{eqnarray}
\delta q^{NS }(z,Q_0^{2})= \delta q^{S }(z,Q_0^{2})= \delta(z -
1)\\
\delta g(z,Q_0^{2})=0
\end{eqnarray}

Thus their moments are

\begin{eqnarray}
\delta q^{NS}(n,Q_0^{2})= \delta q^{S }(n,Q_0^{2})= \int^{1}_{0}
z^{n-1} \delta(z - 1)dz=1\\
\delta g(n,Q_0^{2})=0
\end{eqnarray}

 In the valon model, the hadron structure is obtained by the convolution of
valon structure and  its distribution inside the hadron. Having
specified the various components that contribute to the spin of a
valon, we now turn to the polarized hadron structure, which is
obtained by a convolution integral as follows:

\begin{equation}
g^{h}_{1}(x,Q^{2})=\sum_{\it{valon}}\int_{x}^{1}\frac{dy}{y}
\delta G_{\it{valon}}^{h}(y)
g^{\it{valon}}_{1}(\frac{x}{y},Q^{2}).
\end{equation}

The valon structure is generated by perturbative dressing in QCD.
In such processes with massless quarks, helicity is conserved and
therefore, the hard gluons can not induce sea quark polarization
perturbatively. According to this description, it turns out that
sea polarization is consistent with zero \cite{7}. This finding is
supported by  HERMES experiment and by released data from COMPASS
experiment \cite{12,13,14,15}. Therefore  we have no sea
polarization in our model.

Using the initial conditions in Eq. (9) and Eq. (10), the
calculation of the PPDFs inside the valon  follow from the
standard DGLAP evolution equations.
 The algorithm for calculation of the PPDFs inside the proton can be
decomposed in the following three steps:

\begin{itemize}
\item Calculating PPDFs in the valon by using the DGLAP equations;
\item With a phenomenological approach, one can find the
helicity distributions for the valons, Figure 2. These functions
are $Q^2$-independent. Since we find the valon helicity
distributions, one can use them to calculate  the polarized
nucleon structure up to  $Q^2=10^7 GeV^2$;
\item By using the convolution integral (equations (3) and
(11)), one finds  PPDFs in the nucleon and polarized
 structure function;
\end{itemize}

There is an  excellent agreement between the model predictions
with the experimental data for spin structure functions. A sample
is given in Figure 3.

\begin{figure}[htp]
\centerline{\begin{tabular}{cc}
\includegraphics[width=8cm]{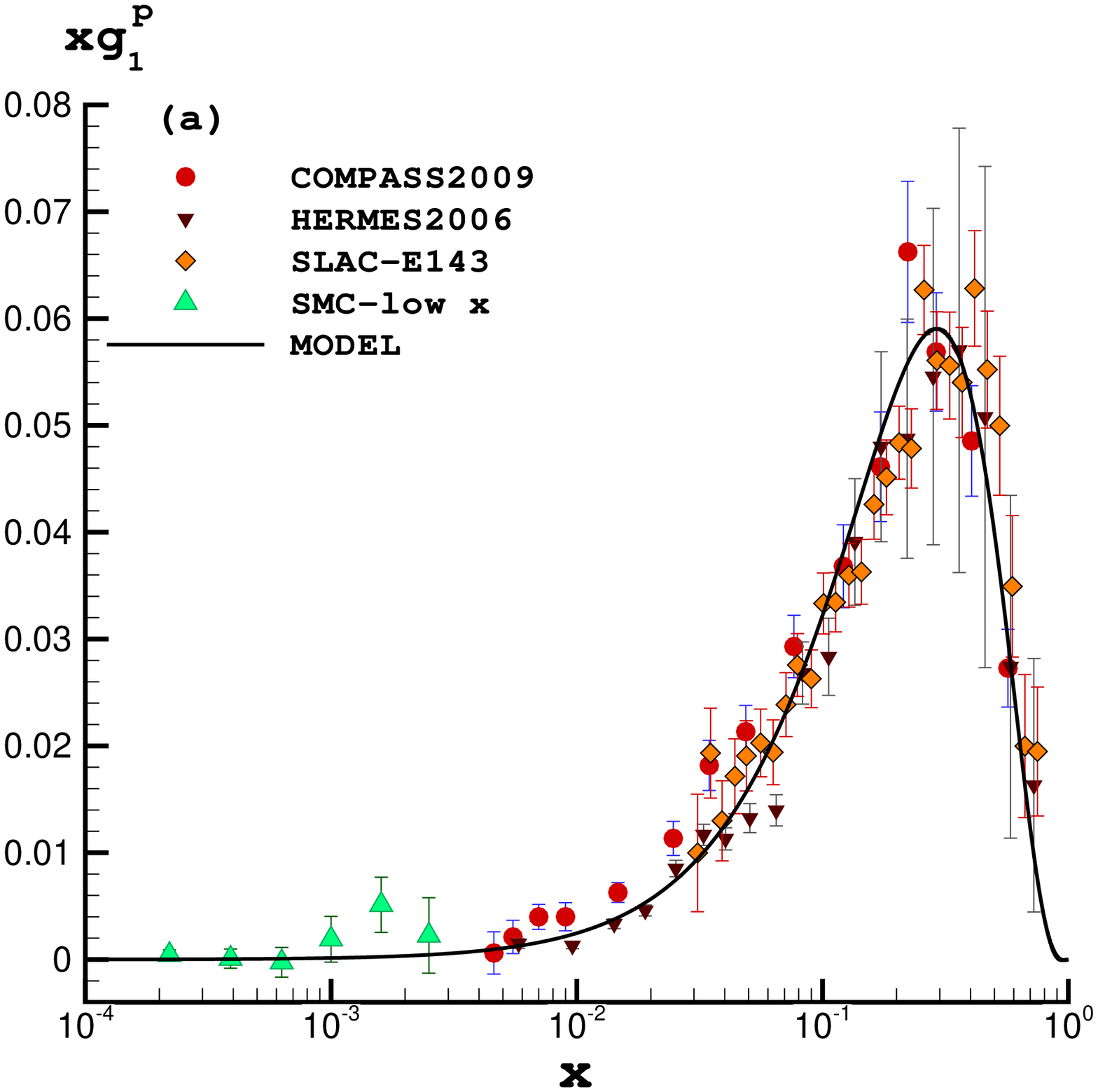}
\includegraphics[width=8 cm]{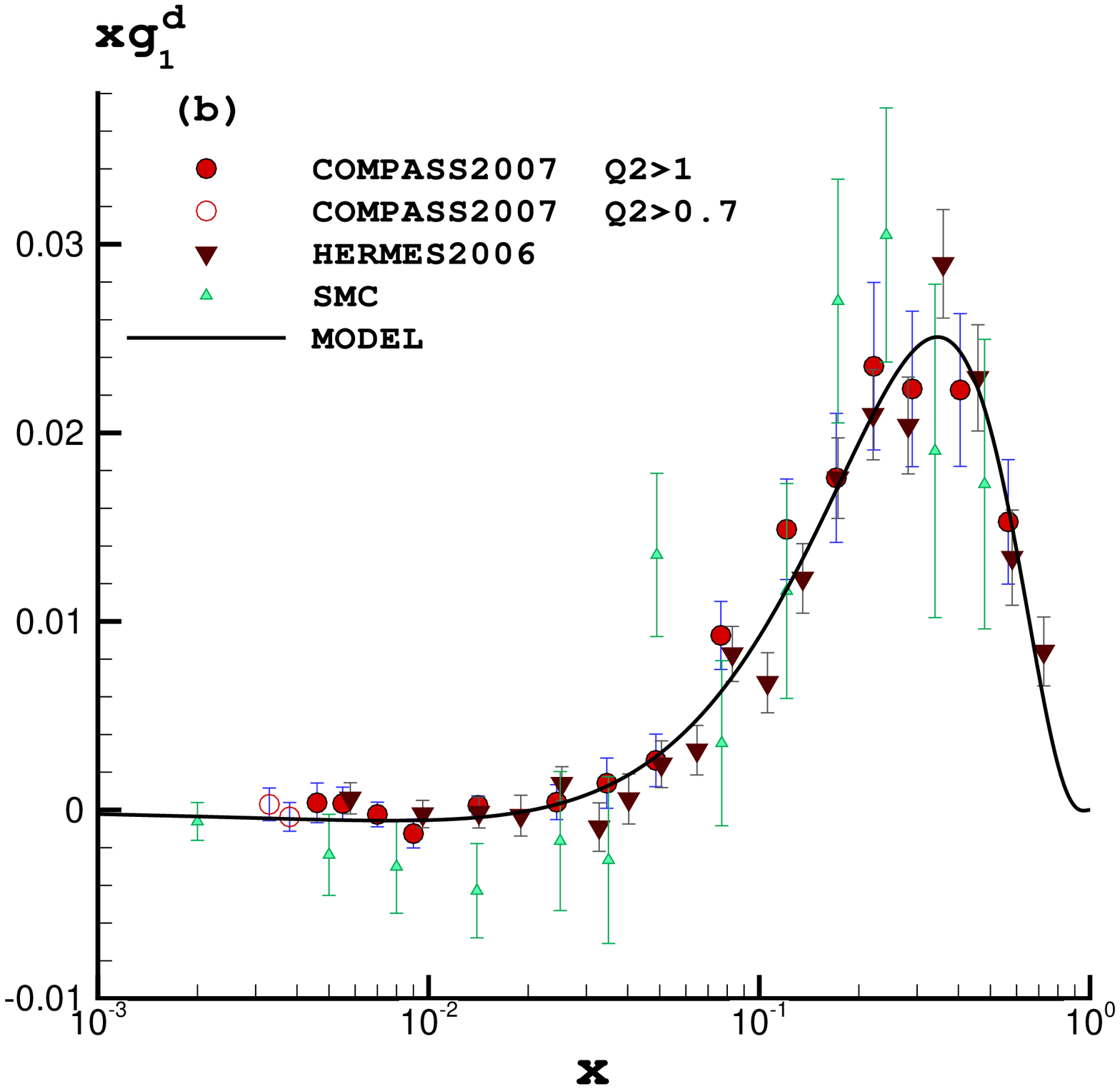}
\end{tabular}}
 \caption{\footnotesize (Color online)  (a) Polarized proton
structure function, $x g_{1}^{p}$ at $Q^2=5 GeV^2$
($\frac{\chi^2}{N}=1.7$). (b) Polarized deuteron structure
function, $x g_{1}^{d}$ at $Q^2=3 GeV^2$
($\frac{\chi^2}{N}=1.33$). The results from model \cite{7} are
compared with the experimental data \cite{9,12,14,16,17,18}. The
data from \cite{9} are newly released data from COMPASS. }

\end{figure}

\section{Non-Singlet spin structure function  }

The non-singlet polarized parton distribution function is defined
as:
 $\Delta q^{NS} (x,t) = (\Delta u - \Delta d)(x,t)$; with $t = \ln
\frac{Q^2}{\Lambda^2}$ .\\
The DGLAP equation for $\Delta q^{NS} (x,t)$ is:
\begin{equation}
\frac{d}{dt} \Delta q^{NS} = \int_x^1 \frac{dy}{y} P \biggl(
\frac{x}{y} \biggr) \Delta q^{NS} (y,t)
\end{equation}
where $P (z=x/y)$ is the NLO spin-dependent splitting function
 \cite{10}. As mentioned before, in our
calculations $\Delta q_{sea}\simeq 0$ , therefore  $\Delta q^{NS}
$ becomes $(\Delta u_{V} - \Delta d_{V})(x,t)$ which is shown in
Figure 4 in comparison with other global fits. The non-singlet
spin structure function is defined as:
\begin{equation}
g_{1}^{NS}\equiv
g_{1}^{p}-g_{1}^{n}=2[g_{1}^{p}-g_{1}^{d}/(1-1.5\omega_D)]
\end{equation}
where $\omega_D=0.058$ accounts for the D-state admixture in the
deuteron wave function.

\begin{figure}[htp]
\begin{center}
 \includegraphics[width=8 cm]{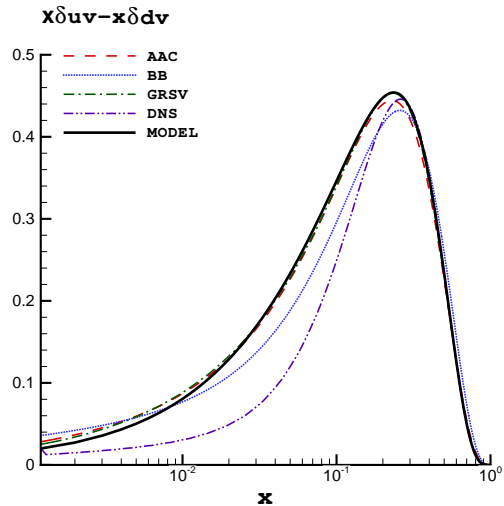}

\caption{\footnotesize (Color online) The non-singlet polarized
parton distribution function in the valon model in comparison with
other global fits \cite{19,20,21,22}. } \label{figure 7.}
\end{center}
\end{figure}

 The $g_1^{NS}$
data is consistent with the quark model and the perturbative QCD
predictions in the valence region $x > 0.2$ \cite{23}.

In Figure 5(a) ,the x dependence of $xg_{1}^{NS}$  is shown in
comparison with data from HERMES, E143, SMC and also with the
newly released data from COMPASS \cite{9}. The results are in very
good agreement with experimental data for the entire measured
range of x.
 In Figure 5(b) The evolution of the Bjorken
integral, derived from the  Figure 5(a),
$\int_{xmin}^{1}(\frac{1}{x}(xg_{1}^{NS})dx=\int_{xmin}^{1}(g_{1}^{NS})dx$
as a function of $x_{min}$ is shown for the model  compared with
the recent HERMES and COMPASS Collaboration data \cite{9,12}. Note
that about 50 percent of the sum rule comes from x values below
about 0.12 and that about 10-20 percent comes from values of x
less than about 0.01. It shows that, $g_{1}^{NS}$ receives a
considerable contribution from the small x region. Thus, it seems
that investigation of small x region of the structure function is
important. In the following
section we will consider $g_{1}^{NS}$  at this region. \\
 In Table 2 we
compared the integral over different x ranges at different scales
of $Q^2$, as
 determined from the valon model with the experimental results
 from COMPASS,
 HERMES, E143, E154, E155, SMC and JLAB also with the recent results from NN
 Collaborations \cite{9,11,16,17,18,24,25,26,27}. We have used data from JLAB for $Q^2>0.5 GeV^2$ to
make sure that the non perturbative effects are small .

\begin{figure}[htp]
\centerline{\begin{tabular}{cc}
\includegraphics[width=8 cm]{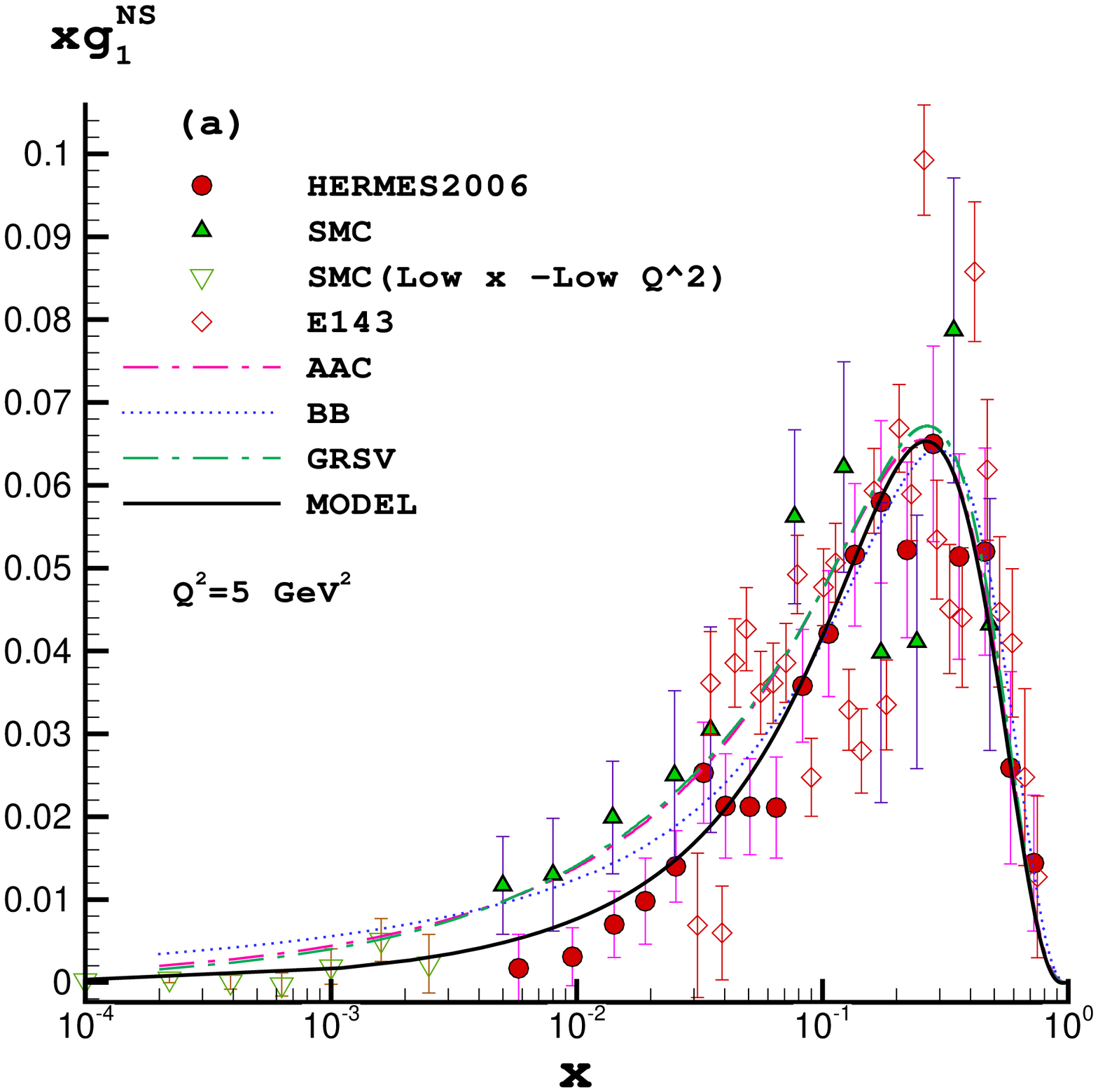}
\includegraphics[width=8 cm]{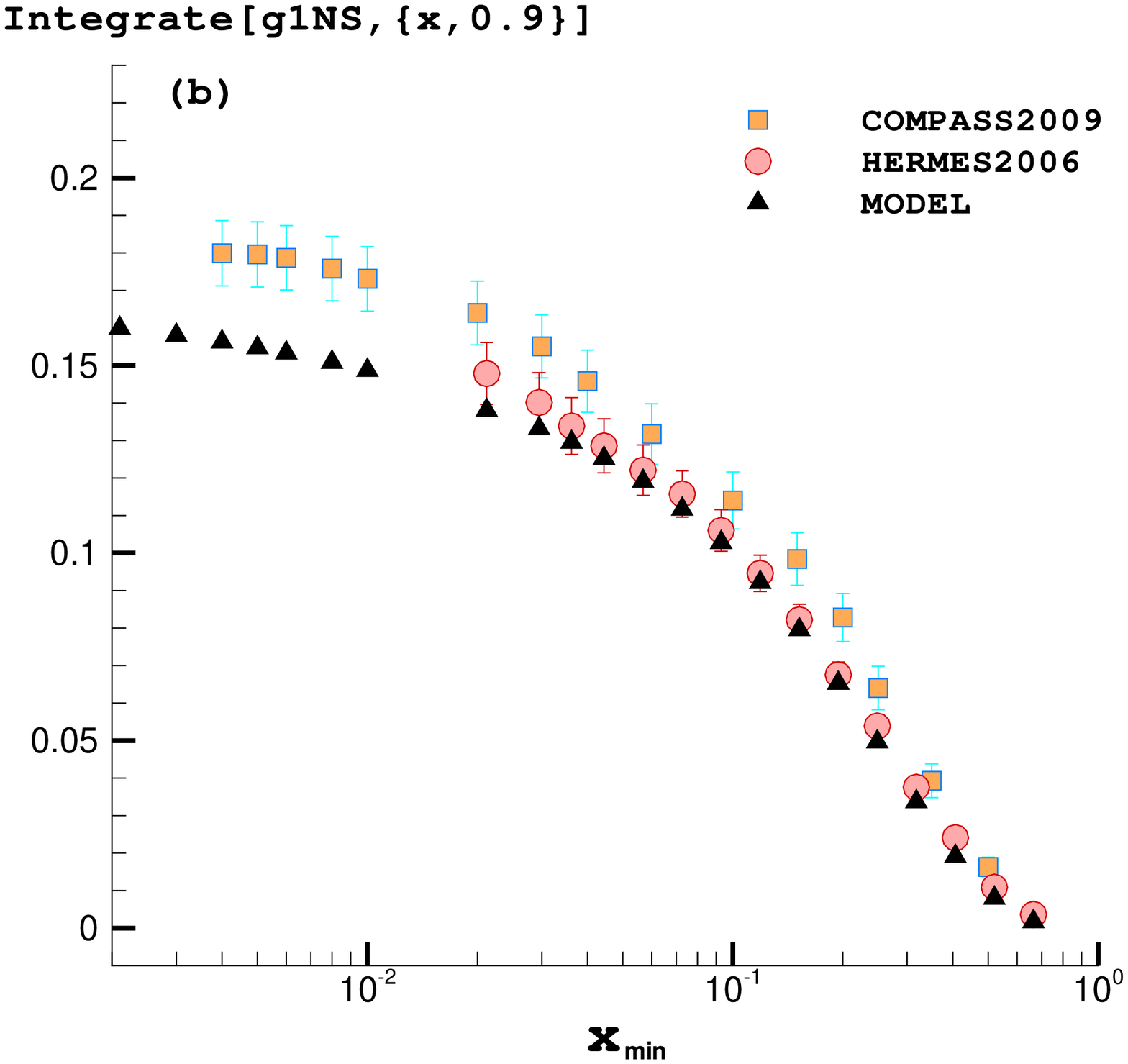}
\end{tabular}}
\caption{\footnotesize (Color online) (a)  $xg_{1}^{NS}$ at $Q^2=5
GeV^2$
 compared with the experimental data
and the results from global fits \cite{19,20,21}
($\frac{\chi^2}{N}=1.25$) . (b)  The integral of $g_{1}^{NS}$ over
the range $0.02 < x < 0.9$ measured by HERMES and COMPASS
Collaborations as  a function of the low $x$ limit of integration,
$x_{min}$, evaluated at $Q^2=10 GeV^2$ in comparison with our
results.} \label{figure 6.}
\end{figure}
\vspace{0.5cm}

\begin{table}
{\footnotesize
\begin{tabular}{|c|c|c|c|c|}
  \hline
  Experiment & $x-Range$  & $Q^2$ & $\Gamma_{1}^{NS}$ & This Analysis \\
  \hline
   \hline
COMPASS &$0.004<x<0.7$  & $3$ & $0.175 \pm0.009\pm0.015$ & $0.1421 $ \\
  \hline
  HERMES &$0.021<x<0.9$  & $5$ & $0.1479\pm0.0169$ & $0.1381 $ \\
  \hline
  E143 & $0.03<x<0.8$  & $2$ & $0.149\pm0.016$ & $0.1276 $ \\
  \hline
  E143 & $0.03<x<0.8$  & $3$ & $0.164\pm0.023$ & $0.1301$ \\
  \hline
  E143 & $0.03<x<0.8$  & $5$ & $0.141\pm0.013$ & $0.1327 $ \\
   \hline
  E154 & $0.03<x<0.8$  & $5$ & $0.168\pm0.010$ & $0.1327 $ \\
   \hline
  E155 & $0.03<x<0.8$  & $5$ & $0.176\pm0.008$ & $0.1327 $ \\
  \hline
  SMC & $0<x<1$  & $10$ & $0.198\pm0.023$ & $0.1626$ \\
  \hline
  SMC & $0<x<1$  & $5$ & $0.174\pm0.024\pm0.012$ & $0.1569 $ \\
   \hline
   JLAB & $0.001 <x<0.8$  & $0.592$ & $0.1027\pm0.0228\pm0.0052$ & $0.123 $ \\
 \hline
 JLAB & $0.001 <x<0.8$  & $0.707$ & $0.0945\pm0.0201\pm0.0151$ & $0.1298 $ \\
 \hline
  JLAB & $0.001 <x<0.8$  & $0.844$ & $0.1021\pm0.0193\pm0.0174$ & $0.134$ \\
 \hline
  JLAB & $0.001 <x<0.8$  & $1.20$ & $0.1307\pm0.0192\pm0.0145$ & $0.1323$ \\
 \hline
  JLAB & $0.001 <x<0.8$  & $1.44$ & $0.1522\pm0.0186\pm0.0089$ & $0.1433$ \\
 \hline
  JLAB & $0.001 <x<0.8$  & $1.71$ & $0.1605\pm0.0182\pm0.0069$ & $0.1318 $ \\
 \hline
 JLAB & $0.001 <x<0.8$  & $2.05$ & $0.1678\pm0.0177\pm0.0056$ & $0.1475 $ \\
 \hline
 JLAB & $0.001 <x<0.8$  & $2.44$ & $0.1666\pm0.0167\pm0.0045$ & $0.1492 $ \\
 \hline
 JLAB & $0.001 <x<0.8$  & $2.92$ & $0.1789\pm0.0106\pm0.0035$ & $0.1511$ \\
 \hline
NN Collaboration & $0.021<x<0.9$  & $5$ & $0.1315\pm0.0144$ & $0.1381$ \\
  \hline
\end{tabular}
\caption{\label{label}Comparison of the integral over different x
range at different scale of $Q^2$, as
 determined from the valon model, with the experimental results
 from COMPASS,
 HERMES, E143, E154, E155, SMC and JLAB also with the results from NN
 Collaborations.}}
\end{table}

\newpage
\section{Regge behavior of  $g_{1}^{NS} $ and full $g_{1}^{p} $ at small x }
In all the results from experimental data for unpolarized and
polarized structure functions, it is seen that these structure
functions increase when x decreases and $Q^2$ increases for fixed
values of x and $Q^2$ respectively. The small x behavior of spin
dependent structure functions reflects the high energy behavior of
the polarized virtual compton scattering total cross section with
increasing total CM energy squared $W^2$ since $W^2=Q^2 (\frac
{1}{x}-1)$. When $W^2>>Q^2$, x is small and $W^2\approx Q^2/x$ and
then at this region the structure functions have scaling behavior.
This is, by definition, the Regge limit and so the Regge pole
exchange picture is therefor quite appropriate for the theoretical
description of this behavior  \cite{28}. The small x or high
energy behavior of the spin structure function is an important
issue  for the extrapolation of data needed to test spin sum rules
for the first moment of $g_{1}$. The small x measurements, besides
reducing the error on the first moment, would provide valuable
information about Regge and QCD dynamics at small x where the
shape of $g_{1}$ is particularly sensitive to the different
theoretical inputs.\\In the case of unpolarized structure
function, $F_{2}$, it is believed that a Regge trajectory combined
with a soft and a hard pomeron can accurately represent the
experimental data \cite{29}. It is
interesting to investigate this issue in the polarized case.\\
The Regge pole model gives the following small x behavior  of the
structure functions $g_1^i(x,Q^2)$\cite{28}
\begin{equation}
g_1^i(x,Q^2)=\gamma_i(Q^2) x^{-\alpha_{i}}
\end{equation}
where $g_1^i(x,Q^2)$ denote either singlet
($g_1^s(x,Q^2)=g_1^p(x,Q^2)+g_1^n(x,Q^2)$) or non-singlet
($g_1^{ns}(x,Q^2)=g_1^p(x,Q^2)-g_1^n(x,Q^2)$) combination of
structure functions. Therefore we expect that describe the small x
behavior of  $g_1^{NS}$  and $g_1^{p}$ structure functions with
one and two Regge exponents respectively. It appears that the
present $g_{1}^{NS} $ data for available small x in the interval
$0.0001<x<0.01$ can be described with a single Regge type exchange
as :
\begin{equation}
g_{1}^{NS}\equiv g_{1}^{p}-g_{1}^{n}  \simeq A x^{\alpha_{Regge}}
\end{equation}

The Regge intercept which governs the small x physics  has smooth
$Q^2$ dependence, in which case one would see $g_{1}^{NS} $ rising
like $x^{\alpha} $ where $-0.5\leq\alpha\leq0$ also at low $Q^2$
and in the measured "small x" region \cite{30} . This value
 varies from  -0.13 to -0.3 when  $Q^2$ is moved from $Q^2=2 GeV^2$ to $Q^2=10
GeV^2$ in the valon model.
 According to the results of Ref.\cite{31},
asymptotic scaling of $g_{1}^{NS} $ depends on one variable
$Q^2/x^2$   only, instead of two variables x and $Q^2$ with the
constant intercept equal to 0.42:
\begin{equation}
g_{1}^{NS}  \simeq (Q^2/x^2)^{\Delta_{NS}/2} \simeq
Q^{\Delta_{NS}} x^{-\Delta_{ NS}} ;\Delta_{ NS}=0.42
\end{equation}

However, it is valid for a very small x only. The applicability
region of that analysis is $x\leq10^{-6}$ . Figure 6  shows the
non-singlet spin structure function at $Q^2=5 GeV^2$ for small x
region. In Figure 7, we fit the non-singlet spin structure
function at small x ($0.0001<x<0.01$) and find the associated
$\alpha_{Regge}$ :

\begin{equation}
g_{1}^{NS} = A x^{\alpha_{Regge}} ;A=0.173 , \alpha_{Regge}=-0.323
\end{equation}

\begin{figure}[htp]
\begin{center}
 \includegraphics[width=8 cm]{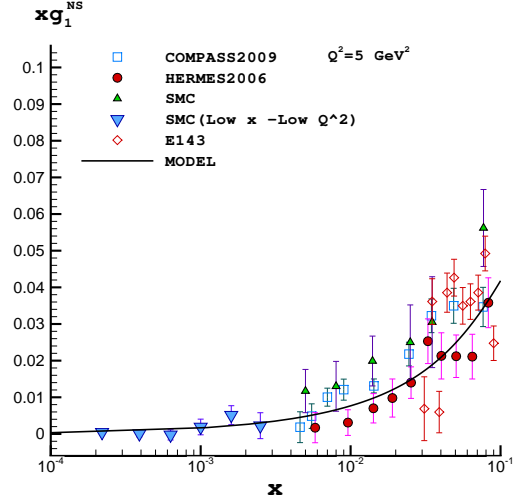}

\caption{\footnotesize (Color online) Small x behavior of
$xg_{1}^{NS} $ at $Q^2=5 GeV^2$  calculated by using the valon
model in comparison with the experimental data . } \label{figure
4.}
\end{center}
\end{figure}

\begin{figure}[htp]
\begin{center}
 \includegraphics[width=8 cm]{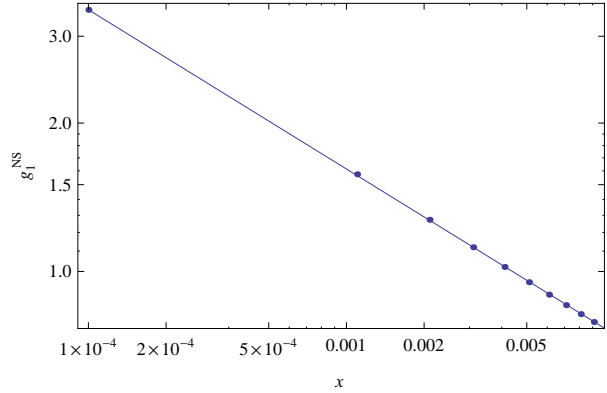}

\caption{\footnotesize  Small x behavior of $g_{1}^{NS} $ at
$Q^2=5 GeV^2$  calculated by using the valon model and using the
best fit to  calculate the Regge exponent. Data points are from
the model and the goodness of fit is: $\chi^2=0.99$. }
\label{figure 5.}
\end{center}
\end{figure}

\subsection{Scaling behavior of  $g_{1}^{p} $ at small x }
Having analyzed the small x behavior of $g_{1}^{NS} $, it is
interesting to see how   $g_{1}^{p} $   behaves as $x
\rightarrow0$. Because the valon model has very good agreement
with existent small x data, it is a good candidate to show the
small x behavior of $g_1^p$ at small x. Actually the results for
this scaling behavior should be compared with small x
data for $g_1^p$. \\
We show that the polarized proton structure function has this
scaling behavior for $1.2<Q^2(GeV^2)<100$ at small x
($10^{-5}<x<10^{-2}$)

\begin{equation}
g_{1}^{p}(x,Q^2)=\sum_{i=1}^{2} a_if_i(Q^2)x^{\varepsilon _i}
\end{equation}

Where $a_i$ and $\varepsilon _i$ are constants and the functions
$f_i(Q^2)$ have this simple general form

\begin{eqnarray}
 f_2(Q^2)=(\frac {Q^4}{1+Q_0^4})^{D_i}\\
 f_1(Q^2)=f_2(Q^2) g(Q^2)\\
\end{eqnarray}

Where
\begin{equation}
g(Q^2)=g_0 +g_1 Log(Q^2)+g_2 Log(Q^2)^2+g_3 Log(Q^2)^3\\
\end{equation}

The results for parametrization of $g_1^p$ are summarized in table
3. (We should noted that if we try to do a fit with only one regge
exponent and only with the first term in Eq. 18, we give a fit
with $\chi^2=0.99$. So, It is clear that we can  describe the
polarized proton structure function with two exponents,
$\chi^2=0.996$,  better than one exponent )

\begin{figure}[htp]
\begin{center}
 \includegraphics[width=8 cm]{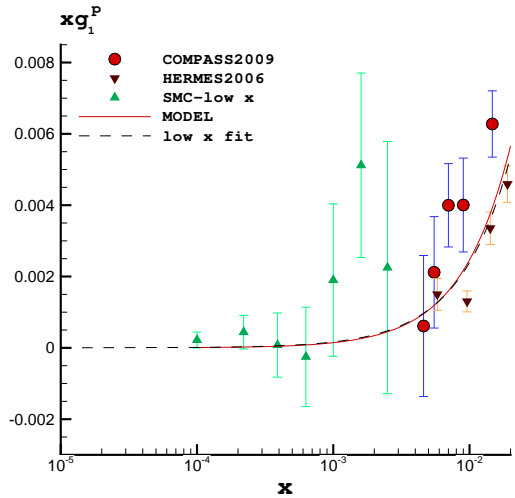}

\caption{\footnotesize (Color online) Small x behavior of  $x
g_{1}^{p} $ at $Q^2=5 GeV^2$  calculated by using the valon model
and using the best fit to  calculate the scaling exponents at
small x . The experimental data selected for a various of $Q^2$ at
small x region.} \label{figure 5.}
\end{center}
\end{figure}

\begin{table}
\begin{center}
\begin{tabular}{|c|c|}
  \hline
  $Parameters$ & $values$   \\
  \hline
   \hline
 $\varepsilon_1$ & 0.196  \\
   \hline
 $\varepsilon_2$ & 0.094  \\
   \hline
  $a_1$ & 0.0215  \\
  \hline
  $a_2$ & 0.0513 \\
  \hline
    $D_1$ &0.759 \\
   \hline
 $D_2$ & 0.434  \\
    \hline
$g_0$ & 17.538 \\
    \hline
 $g_1$ & -11.809  \\
    \hline
$g_2$ & 2.652  \\
    \hline
    $g_3$ & -0.200 \\
    \hline
  $Q_0$ & 1.300   \\
\hline
  \hline
$\chi^2$(Goodness of fit) & 0.996  \\
  \hline
\end{tabular}
\caption{\label{label} global fit parameters obtained by fitting
the Eq. (4.4) with the  data extracted from the valon model at
small x.}
 \end{center}
\end{table}

In Figure 8 we show the Small x behavior of  $x g_{1}^{p} $ at
$Q^2=5 GeV^2$  calculated by using the valon model and using the
best fit to  calculate the scaling exponents, $\varepsilon_i$.\\
As a result, we concluded that only by using two scaling
exponents, we can describe the small x behavior of $g_1^p(x,Q^2)$
well. This behavior has also been seen in the small x tail of
$F_2^p(x,Q^2)$ structure function \cite{28}. The question has been
raised  \cite{32,33}  whether the observed rise on
$g_1^{NS}(x,Q^2)$ follows from one or two pomeron exchange( a
polarized analogue of the one or two pomerons question). However,
looking at $g_1^p(x,Q^2)$, it indeed requires two pomerons!

\section{Conclusion}
In this paper we calculated the non-singlet spin structure
function, $g_{1}^{NS} $, of the nucleon in the  valon model.  The
results of these calculations are in excellent agreement with all
experimental data for the entire measured range of x. We also
study the small x behavior of non-singlet spin structure function
and the Regge behavior of $g_{1}^{NS}$ to calculate the Regge
exponent, $\alpha_{Regge}$. Finally,  we studied the scaling
behavior of $g_1^p$ at small x. We conclude that only by using two
scaling exponents, we can describe the small x behavior of $g_1^p$
 well. This is very similar to existence of two soft and hard
pomeron to describe the  small x behavior of $F_2(x,Q^2)$ in
unpolarized case. It is shown that the valon model can predict the
polarized nucleon structure functions for the entire measured x
range very well. The validity range for using the valon model is
$0.5<Q^2(GeV^2)<10^7$ and $10^{-6}<x<1$. In this model
 sea quarks polarization is consistent
  with zero (the finding that was confirmed by very recent experiments at HERMES and COMPASS),
    so the polarized hadron structure functions can build only by
  polarized valence and gluon distribution by finding only one
 type of polarized valon distribution for each kind of valons.
   Between $0.3<Q^2(GeV^2)<0.5$ we should
consider other effects such as combination of resonance physics
and vector-meson dominance that are important at low $Q^2$. Chiral
perturbation theory may describe the behavior of polarized nucleon
structure close to threshold. These issues will be considered in
our future works to improve our model.

\section*{Acknowledgments}
 We would like to thank Professor G.Altarelli for
his careful reading of the manuscript and  for the productive
discussions. F.Taghvi-Shahri thanks  A.Korzenev for his help and
also  thanks  Dr M.Khakzad for reading the manuscript.

\end{document}